# Mechano-freezing of the ambient water

Xi Zhang,[1,2,3a] Tingting Yan,[1,a] Bo Zou[1], and Chang Q Sun[1,2,3]

[1] State Key Laboratory of Superhard Materials, Jilin University, Changchun 130012, China

[3] *NOVITAS, School of Electrical and Electronic Engineering, Nanyang Technological University, Singapore 639798*

[2]*Center for Coordination Bond and Electronic Engineering, College of Materials Science and Engineering, China Jiliang University, Hangzhou 310018, China*

Ecqsun@ntu.edu.sg

[a]X.Z. and T.Y. contribute equally.

Abstract

Raman spectroscopy examination of the 25°C water freezing under compression revealed transition from 1.35 GPa to 0.86 GPa upon ice being formed at continued volume change. The transition is associated with a slight blue shift of the high-frequency phonon ($\omega_H$ ~ 3120 cm$^{-1}$) and creation of the low-frequency phonons ($\omega_L \le 200$ cm$^{-1}$). In the liquid and in the solid phase, the increased pressure softens the $\omega_H$ and stiffens the $\omega_L$, which indicates the presence of the inter-electron-pair repulsion in both liquid and solid water.

**Phonon frequencies.** The vibrational frequency shift, probed using FTIR or Raman spectroscopy, is proportional to the square root of bond stiffness, approximates directly the length and strength change of the bond during relaxation. Comparing the energy of a vibration system to the Taylor series of the inter-atomic potential energy, $u_x(r)$, leads to the dimensionality of the vibration (phonon) frequencies for an uncoupled system [1]:

$$\Delta\omega_x \propto \left( \left. \frac{\partial^2 u_x(r)}{\mu \partial r^2} \right|_{r=d_x} \right)^{1/2} \propto \sqrt{E_x/\mu}\Big/ d_x \propto \sqrt{Y_x d_x}$$

(1)

The stiffness is the product of the Young's modulus ($Y_x \propto E_x/d_x^3$) and the length of the segment in question [2]. The $\mu$ is the reduced mass of the vibrating dimer. Therefore, the frequency shift measures directly the segmental stiffness based on the dimensionality analysis.

However, the Coulomb coupling in the O:H-O bond revises the phonon dispersion of water and ice [3],

$$\Delta\omega_x \propto \left( \left( k_x + k_c \right) / \mu_x \right)^{1/2} \tag{2}$$

(x= L and H correspond to the O:H and the H-O bond, respectively). where $k_c$ being the second differential of the Coulomb repulsion between the electron pairs on adjacent oxygen atoms [3].

**H-bond.** Figure 1 inset shows the hydrogen (H) bond model with ultra-short-range exchange (H-O) and van der Waals (O:H) interactions. Repulsion between the electron pairs on adjacent oxygen atoms dislocates oxygen atoms in the same direction but different amounts under external stimulus. The segmented H-bond forms a pair of asymmetric, coupled, and H-bridged oscillators whose relaxation in length and energy determines the unusual behavior of water and ice [1, 2, 4].

**FTIR (absorption) and Raman (reflection) phonon spectra of liquid water.** Figure 1 shows the typical FTIR absorption and Raman reflection spectra of water at the ambient. There four features corresponding to, from high to low frequencies: i) the H-O stretching phonons in the skin ($\omega_H$ ~ 3450 $cm^{-1}$) and in the bulk (~ 3200 $cm^{-1}$); ii) the liberation mode of ∠H-O-H bending at $\omega_{B1}$ 1600 ~ 1750 $cm^{-1}$; iii) the O:H-O bending mode at $\omega_{B2}$ ~ 500 $cm^{-1}$ and, iv) the O:H stretching mode at $\omega_L$ ~ 200 $cm^{-1}$. The liberation mode is insensitive to the experimental conditions. The intensity of the peaks below 500 $cm^{-1}$ is rather weak but very sensitive. Measuring both the high-frequency $\omega_H$ and the low-frequency $\omega_L$ would be sufficiently comprehensive for examining the variation of stiffness of the respective H-bond segment.

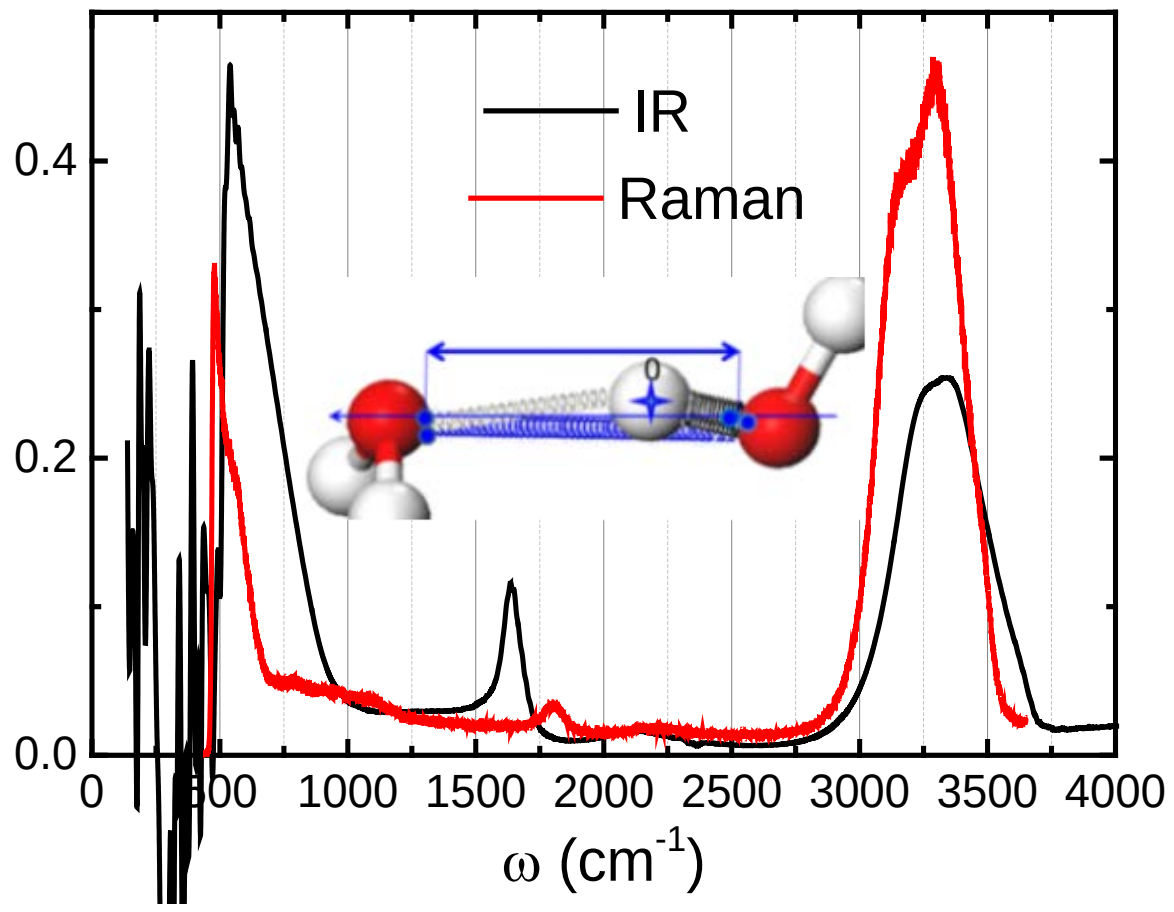

Figure 1 Characteristic spectral peaks of the IR absorption and the Raman reflection from water at the ambient conditions. Peaks correspond to: O:H stretching mode: $\omega_L < 300$ cm$^{-1}$, ∠O:H-O bending mode: $\omega \sim 500$ cm$^{-1}$, ∠H-O-H bending liberation mode: $1500 < \omega < 1800$ cm$^{-1}$. H-O stretching mode in bulk water: $\omega_H \sim 3200$ cm$^{-1}$, H-O stretching mode in the liquid skin: $\omega_H \sim 3450$ cm$^{-1}$. The liberation mode is insensitive to the external stimulus. The inset shows the segmented H-bond model with H as coordination origin.

**Raman spectra of water under compression.** Raman spectra of water under high pressure were measured in a symmetric diamond anvil cell (DAC). The cell consists of two culet diamonds with a face of 1000 μm in diameter. A T301 steel gasket was preindented to 80 μm in thickness, and then, a center hole with a diameter of 500μm was drilled as the sample chamber. Deionized water was placed in the gasket hole together with a small ruby ball to determine the pressure using the standard ruby-fluorescence method.

Figure 2 shows the Raman spectra of 25 °C water as a function of pressure. During the phase transition, the pressure drops from 1.35 to 0.86 GPa while the volume of the diamond cell change continually [5]. The transition is associated with a slight blue shift of the high-frequency phonon ($\omega_H \sim 3120$ cm$^{-1}$) and creation of the low-frequency phonons ($\omega_L \leq 200$ cm$^{-1}$).

In the liquid and in the solid phase, the increased pressure softens the $\omega_H$ and stiffens the $\omega_L$, which indicates the presence of the inter-electron-pair repulsion in both liquid and solid water, which is the same to that happened to ice at 80 K [2]. The sudden drop in pressure upon icing may indicate a new mechanism for the O:H and H-O energy transition. According to water phase diagram, Ice-six (ice VI) should be formed from liquid water at room temperature by increasing the pressure to the range of 0.6 GPa and 2.1GPa.

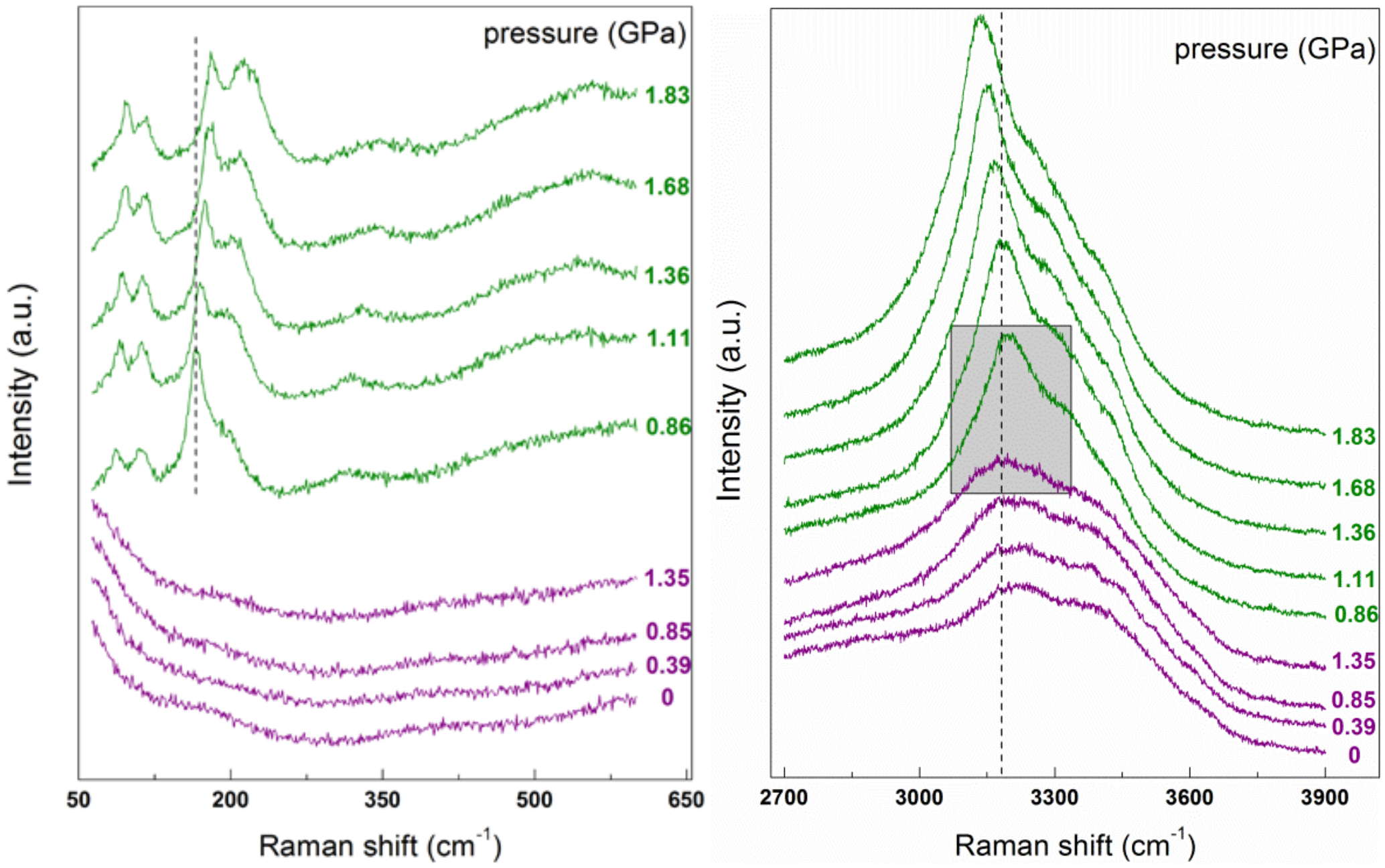

Figure 2 Raman spectra of the mechano-freezing of water at room temperature, showing the pressure drop from 1.35 to 0.86 GPa at freezing while the volume change continuelly [5].

**DFT calculation.**

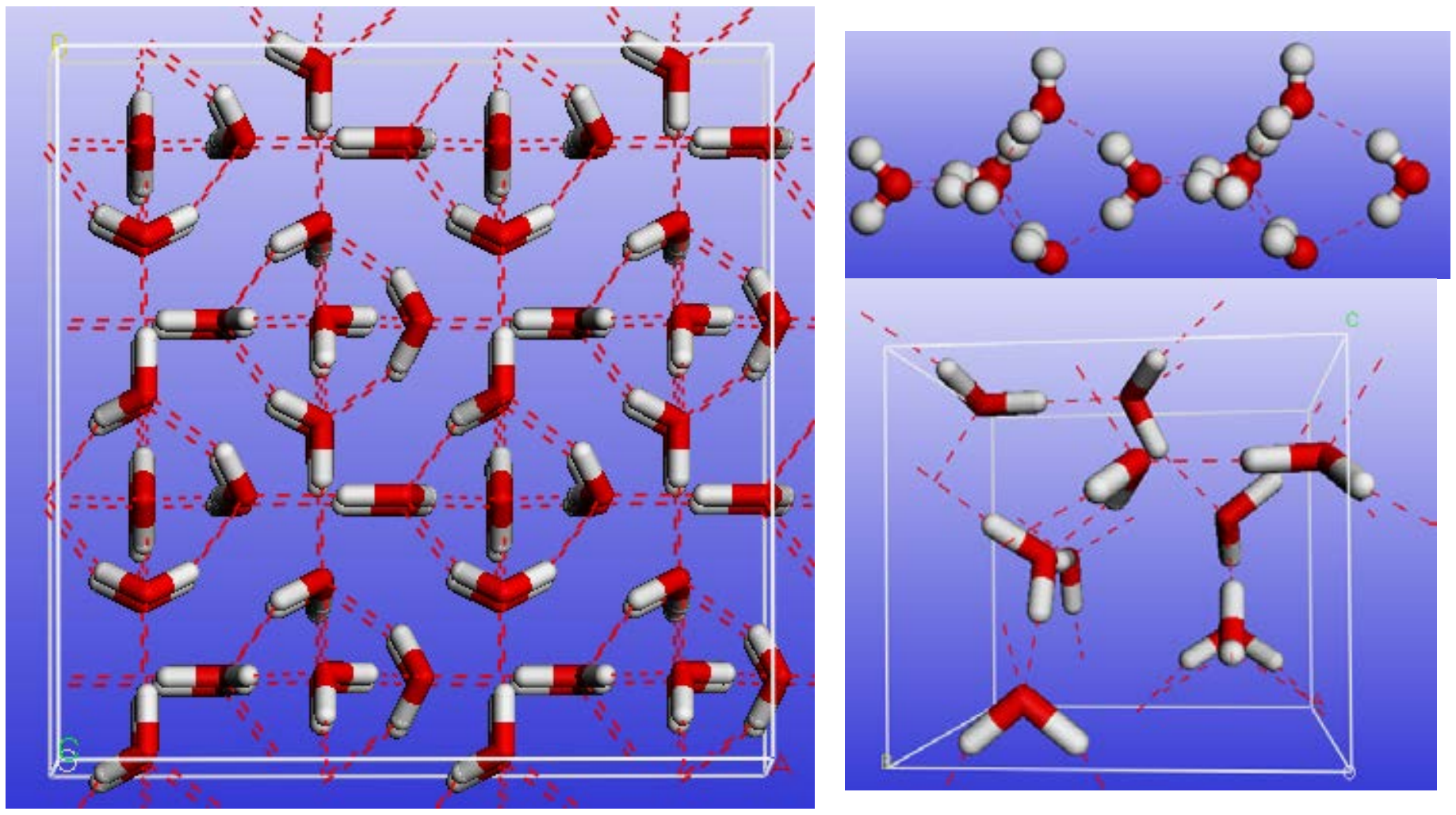

Figure 3 Side-view of ice-VI bulk structure. hydrogen bond network was connected in the form of cage-hexamers. Each network is linked together through the four equatorial water molecules in the hexamers. The unit cell of space group *P42/nmc* used in DFT calculation. Similar to other high-pressure phases of ice, ice-VI has two interpenetrating and independent sublattices.

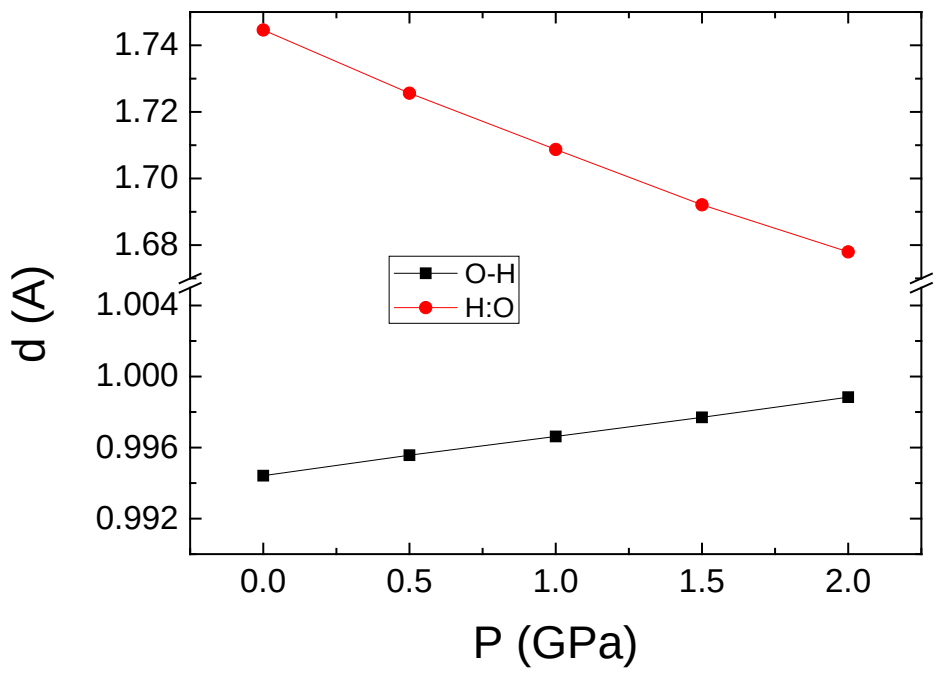

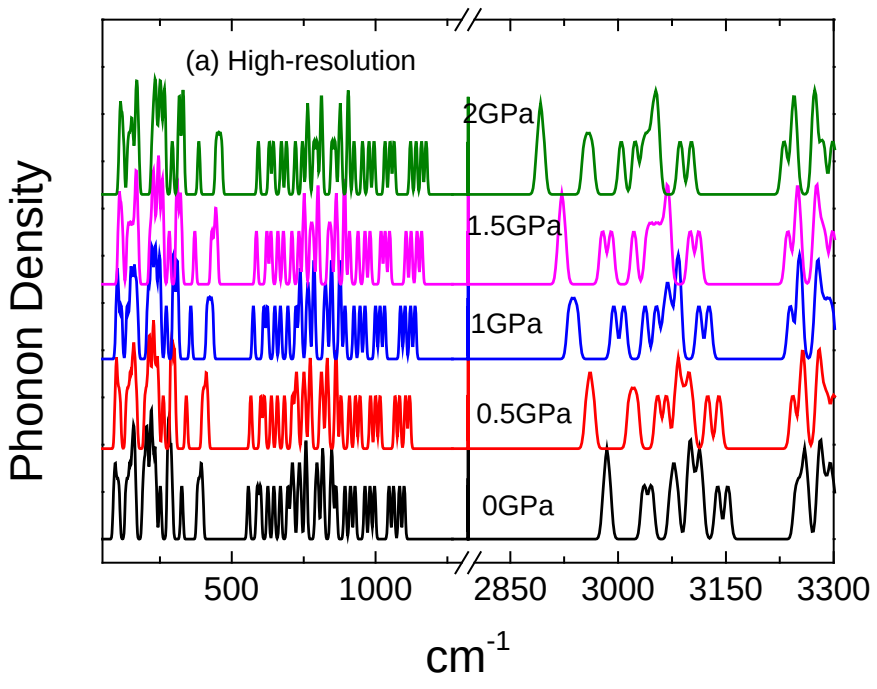

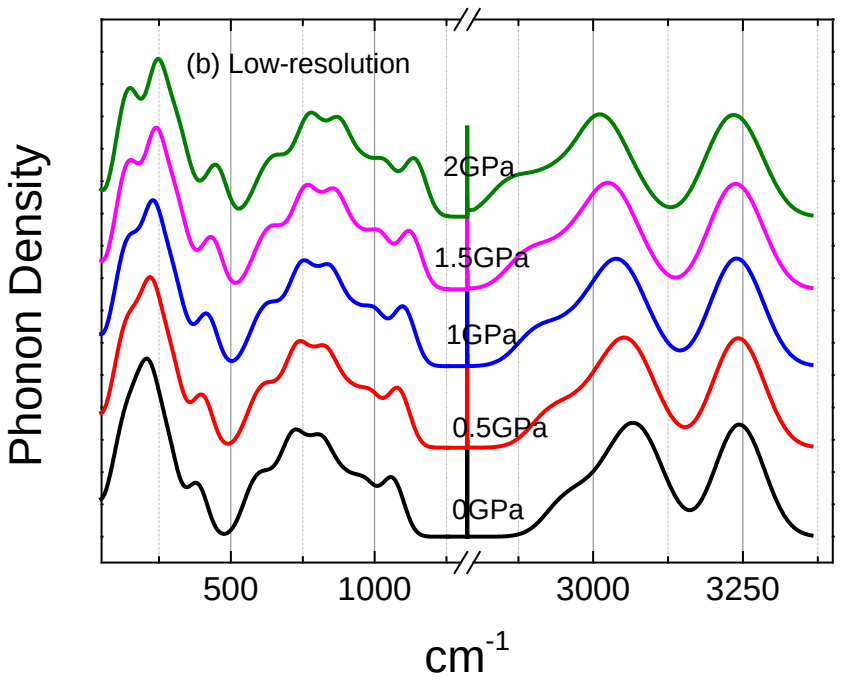

Figure 4 DFT derived cooperative relaxation of the O:H and the H-O bond towards O:H and H-O length symmetrization with an association of the respective $\omega_H$ softening and $\omega_L$ stiffening under compression of ice-VI.

**Summary**

Raman measurements and DFT calculations reveled the cooperative relaxation of the O:H and the H-O bond in water and ice. Liquid-solid phase transition of the ambient water happens at 1.35-0.86 GPa associated with $\omega_H$ stiffening and $\omega_L$ creation. In the liquid and in the solid phase, pressure induced phonon relaxation follows the same trend of ice at low temperature. Results suggest the persistence of the Coulomb repulsion in water and ice.

**DFT Calculation methods.** The ice-VI structure as the calculation object is shown in Figure 3. The symmetry group of ice-VI is *P42/nmc*, which requires a proton-disordered structure. The primary unit of ice-VI is a tetragonal cell with 10 water molecules. DFT calculations on geometry optimization and vibration properties under different pressures were performed by CASTEP code. Perdew-Becke-Enlser (PBE) functional in generalized gradient approximation (GGA) was employed to describe the exchange-correlation energy with dispersion-force correction in Grimme scheme. Norm-conserving pseudopotential is adopted in line with the phonon calculation. Energy cutoff is 750.0eV with 2×2×3 k-point set.

Vibration spectrum was calculated by linear response theory or density functional perturbation theory (DFPT). DFPT is one of the most popular methods of ab initio calculation of lattice dynamics. Linear response provides an analytical way of computing the second derivative of the total energy with respect to a given perturbation.

The phonon frequency ω and polarization vector $\mathbf{u}_0$ at phonon wave vector $\mathbf{q}$ can be described by the eigenvalue problem:

$$\mathbf{D}(\mathbf{q})\mathbf{u}_0 = M\omega(\mathbf{q})^2\mathbf{u}_0 \quad (1)$$

where M is the effective mass matrix; **D** is the force constant matrix (or hessian matrix). **D**(**q**) can also be represented in reciprocal space as a dynamical matrix:

$$\mathbf{D}(\mathbf{q}) = \frac{1}{N_{\mathbf{R}}}\sum_{\mathbf{R}}\mathbf{D}(\mathbf{R})\exp(-i\mathbf{qR}) \quad (2)$$

Thus, classical equations of motion can be written in the language of dynamical matrices. And the displacement of an atom can be described in the form of plane waves:

$$\mathbf{u}(\mathbf{R},t) = u_0 \exp(i\mathbf{qR} - i\omega t) \quad (3)$$

A popular way of classifying lattice vibrations is based on the relationship between the orientation of the polarization vector, $\mathbf{u_0}$, and the propagation direction, **q**.

The Hessian matrix was calculated by 3*30=90 (3 coordinates times 30 atoms) perturbations to an ice-VI unit cell. The phonon frequencies and polarization vector at □point of **q** was calculated by solving the eigenvalue problem in Eq.(1). Statistically, phonon density distribution was obtained from the frequencies with high or low resolutions.